# Discovery of a large magnetic nonlinear Hall effect in an altermagnet

Lei Han[1,†], Xizhi Fu[2,†], Xiaokang Li[3,†], Cheng Song[1,*], Yuxiang Zhu[1], Hua Bai[1], Ruiyue Chu[1], Jiankun Dai[1], Shixuan Liang[1], Zengwei Zhu[3], Maciej Sawicki[4], Junwei Liu[2,*], Feng Pan[1,*]

[1]Key Laboratory of Advanced Materials (MOE), School of Materials Science and Engineering, Tsinghua University, Beijing 100084, China.

[2]Department of Physics, The Hong Kong University of Science and Technology, Hong Kong 999077, China.

[3]Wuhan National High Magnetic Field Center and School of Physics, Huazhong University of Science and Technology, Wuhan 430074, China.

[4]Institute of Physics, Polish Academy of Sciences, Aleja Lotnikow 32/46, PL-02668 Warsaw, Poland.

*Corresponding author. Email: songcheng@mail.tsinghua.edu.cn; liuj@ust.hk; panf@mail.tsinghua.edu.cn.

[†]These authors contributed equally.

## Abstract

**Since Edwin Hall's groundbreaking discovery of the Hall effect in 1879, magnetism, spin, and quantization have been expanding the scope of Hall effects, continuously driving transformative progress in science and technology. Among them, the recent nonlinear Hall effect (NLHE), where a longitudinal electric field tunes quantum geometry to generate nonlinear Hall voltage, attracts wide attention as a sensitive probe of topological phases across a wide range of materials. Here, we report a new Hall effect member: the magnetic nonlinear Hall effect (MNLHE), characterized by the quadratic Hall conductivity dependence on the magnetic field—rather than the electric field as in traditional NLHE. This finding relies on an altermagnet, $Mn_5Si_3$ thin film, whose alternating-sign Berry curvatures from the altermagnetic spin-splitting effect ensure MNLHE clearly distinguishable from the first-order anomalous Hall effect. The observed quadratic dependence originates from chiral next-nearest-neighbor hopping processes that acquire magnetic-exchange-driven Zeeman energies and Haldane-like chiral flux phases. Remarkably, this MNLHE cannot be fitted by an analytic function, i.e., is non-analytic, as reversing the magnetic field flips the alternating spin-splitting bands and reverses the hopping chirality, which is absent in traditional NLHE. Beyond offering a distinctive transport fingerprint for altermagnet $Mn_5Si_3$ thin film, this MNLHE is large and unsaturated up to 60 T, providing opportunities for pulsed high-field sensing technologies in both fundamental research and engineering applications.**

## Main

The ordinary Hall effect is one of the most classical phenomena[1] in condensed matter physics, and over the past few decades, it is hard to overemphasize the impact of the milestone advancements in new Hall effects on the field.[2-7] Discoveries like the anomalous[7], spin[6], and quantum Hall effect[2] are far more than novel phenomena—they serve as defining tools for uncovering fundamental insights into Fermi surface features, spin-orbit coupling, and topological physics, while driving groundbreaking applications like precise sensing, low-power spintronics, and quantum computing. Recently, nonlinear Hall effect (NLHE)[8,9], i.e., higher-order Hall responses to longitudinal electric field (Fig. 1a), have garnered wide attention as both a sensitive probe of quantum geometry[10] and a promising mechanism for energy harvesting[11]. It has been explored in a broad range of materials like 2D materials[9,12-18], twisted systems[19,20], and Weyl/Dirac semimetals[11,21-24].

Typically, NLHE arise from electric field modulation of quantum geometry, and thus these electric-field-relevant NLHE are referred to as electric NLHE here. Taking the representative electric NLHE due to Berry curvature as an example, as shown in Fig. 1b and 1c, bands connected by time-reversal symmetry (*T*), i.e., *T*-paired spin-valley locking (TSVL)[25], exhibit opposite Berry curvatures (in red and blue), with the black circle marking the original Fermi surface and the red/blue circles indicating the electric-field-tilted Fermi surfaces. Berry curvature dipole[8]/quadruple[26] emerges due to Fermi surface tilting, which generates second-order[9]/third-order[27] electric NLHE. Despite showcasing the elegance of inversion symmetry (*P*) breaking[10], these electric NLHE remain inherently weak, and lock-in techniques are often required to detect the AC higher-order Hall voltage $V_{\mathrm{AC}}$ from AC driven electric field $E_{\mathrm{AC}}$, as shown in Fig. 1a.

The magnetic field can also modulate Berry curvature to generate NLHE, which can be termed as magnetic nonlinear Hall effect (MNLHE), a new class of Hall effect (Fig. 1d), as their mechanisms are essentially distinct from electric NLHE. Crucially, the magnetic field may offer far stronger modulation for generating large nonlinear Hall voltage, as it both induces Zeeman band shifts and tunes the order parameter of magnetic materials. However, MNLHE has remained unreported for long. Note that the second-order Hall resistance can be linearly dependent on both the electric and magnetic fields in some materials, i.e., the bilinear magnetoelectric resistance[28,29]. Such phenomena are relevant to magnetic-field-dependent charge-spin interconversion process instead of magnetic-field-modulated Berry curvature. It is challenging to observe MNLHE due to inherent limitations of traditional magnetic materials. They either exhibit single-sign Berry curvature, as in ferromagnets, where the dominant first-order effects like anomalous Hall effect (AHE) overshadow higher-order MNLHE, or retain *PT* symmetry, as in conventional antiferromagnets, suppressing Berry curvature and inhibiting large MNLHE.

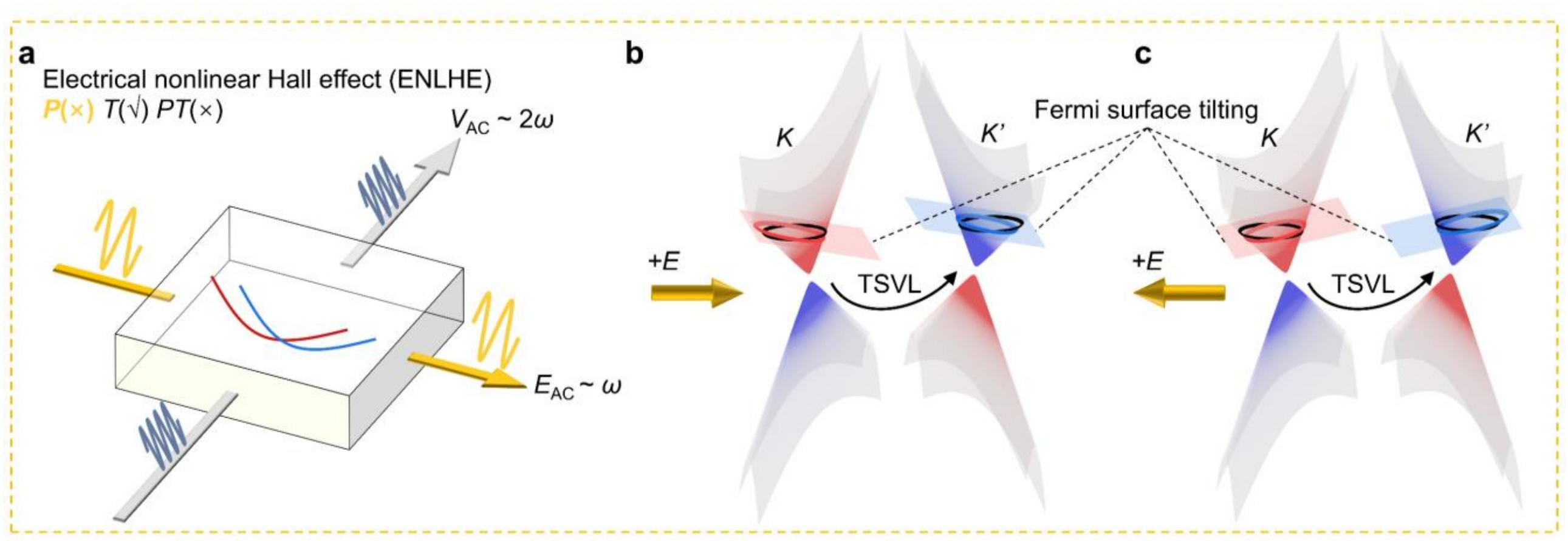

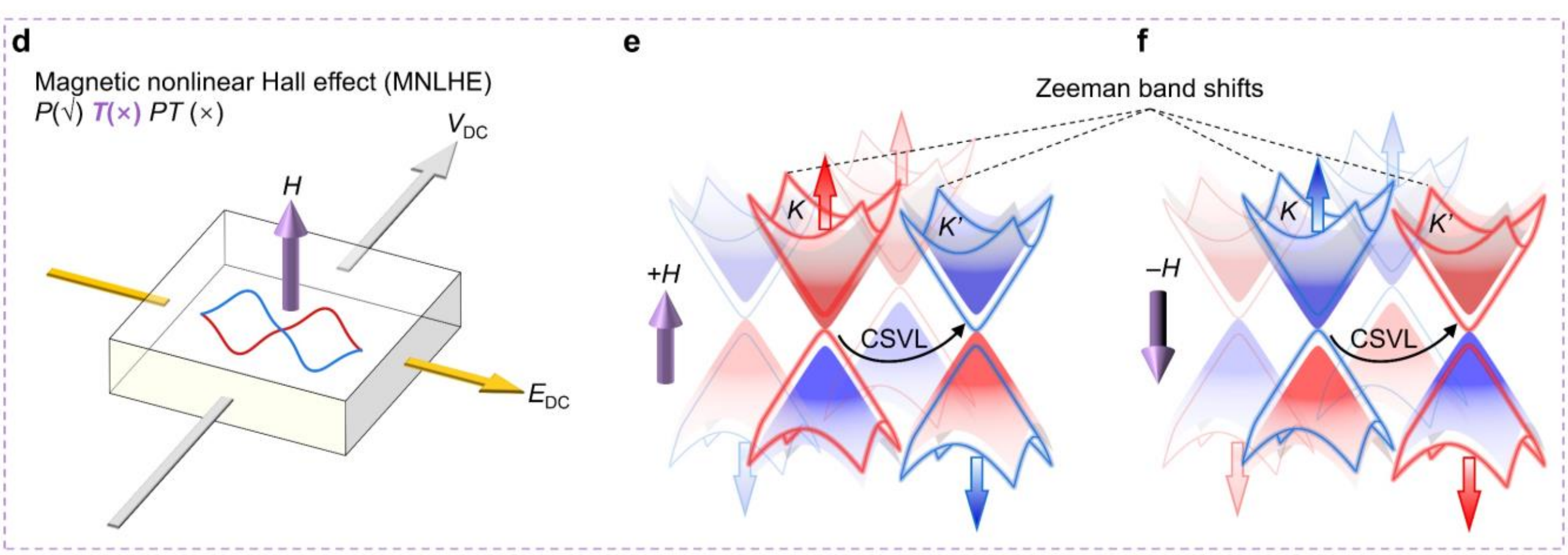

**Figure 1 | Comparison of electric nonlinear Hall effect and magnetic nonlinear Hall effect.** (**a**) Schematic of electric NLHE in materials with *P* symmetry breaking, showing Rashba spin-splitting bands (red and blue lines). Second-order electric NLHE is illustrated here, where a longitudinal AC electric field $E_{AC}$ with frequency $\omega$ generates a nonlinear Hall voltage $V_{AC}$ at $2\omega$, detectable via lock-in amplifiers. (**b, c**) Band, Berry curvature, and Fermi surface modulation under opposite electric fields for electric NLHE. Bands are connected by TSVL, with red and blue indicating opposite Berry curvatures. Black and red/blue circles represent the Fermi surface without and with the electric field modulation, respectively. (**d**) Schematic for MNLHE in altermagnet with *T* symmetry breaking,

showing alternating spin-splitting bands (red and blue lines). DC Hall voltage $V_{DC}$ with nonlinear dependence of the magnetic field is detectable under a DC longitudinal electric field $E_{DC}$. (**e**, **f**) Bands and Berry curvature modulation under opposite magnetic fields for MNLHE in an altermagnet. Bands connected by CSVL of $C_4T$ are shown as an example. Red and blue in the bands represent opposite Berry curvatures, with the color of contours denoting opposite spins. Zeeman shifts of these bands under a magnetic field are shown by the shift of these contours.

Altermagnets[30-32] (AMs) present an ideal platform for observing MNHLEs. Characterized by alternating collinear compensated magnetic moments in real space[33] and alternating spin-splitting bands in momentum space[34], AMs break *PT* symmetry while preserve crystal symmetry-paired spin valley locking (CSVL)[25]. This generates non-zero but alternating-sign Berry curvature $\mathbf{\Omega}$ at CSVL **k**-points ($C\mathbf{\Omega} = -\mathbf{\Omega}$), where $C$ is $C_4T$ as an example shown in Fig. 1e and 1f, denoting the joint operation of fourfold rotational symmetry and time-reversal symmetry. Alternating-sign Berry curvature suppresses strong first-order Hall responses[35-40], as *PT* symmetry breaking in altermagnets is in fact intrinsically tied to higher-order Hall effects[41,42], thereby making higher-order MNHLEs distinguishable. Meanwhile, an external magnetic field can induce a Zeeman shift in the CSVL bands of AMs (Fig. 1e and 1f, denoted by the red and blue contours) to break the $C\mathbf{\Omega} = -\mathbf{\Omega}$ relationship, while simultaneously tunes the order parameter to modulate $\mathbf{\Omega}$, significantly enhancing net $\mathbf{\Omega}$ near the Fermi surface. Therefore, AMs not only create opportunities for large MNHLEs but also, in turn, establish MNHLE as a promising AM's fingerprint by offering deeper insights into alternating spin-splitting bands and Berry curvature multipoles, which represents an urgent but unexplored research direction.

Herein, we report the discovery of a large MNHLE in epitaxially strained $Mn_5Si_3$ thin film, an AM candidate, which can be directly measured by DC Hall voltage and does not saturate even up to ~60 T. The terminology of MNLHE is to emphasize its fundamental difference from electric NLHE. MNLHE is characterized by a quadratic dependence of Hall conductivity on magnetic field due to chiral next-nearest-neighbor hopping, reminiscent of the Haldane model with chiral flux phases. The reversal of hopping chirality under magnetic field produces unique non-analytic feature, which is originated in the flipping of spin-splitting bands and absent in electric NLHE. Beyond its scientific significance, the large and unsaturated MNLHE is highly promising for high-field sensing in applications such as plasma science, condensed matter physics, magnetic flux compression, or magnetic pulse welding.

**Experimental observation of non-analytic quadratic Hall conductivity with magnetic field**

$Mn_5Si_3$ thin films are chosen due to their paramagnet (PM)-to-AM and AM-to-noncollinear AFM (ncAFM) phase transitions, enabling *in-situ* comparisons of Hall responses in these phases.[37,38,43] To be shown later, these comparisons allow us to effectively exclude possible nonlinearity in Hall response due to multi-carrier effect, like in CrSb.[44-46] In each $Mn_5Si_3$ unit cell, four $Mn_a$ and six $Mn_b$ occupy two inequivalent Wyckoff positions (Fig. 2a). As the temperature drops into the AM phase, G-type AFM ordering develops on four of the six $Mn_b$ atoms in two atomic layers at $z = ¼$ and $z = ¾$ (green planes in Fig. 2a), forming the Néel vector $N$ as the order parameter. In bulk $Mn_5Si_3$, the PM-to-AM magnetic phase transition is accompanied by the hexagonal-to-orthorhombic structural transition, with the resulting orthorhombic structure conserving $PT$ and $tT$ symmetry.[47,48] In contrast, for $Mn_5Si_3$ thin film, this hexagonal-to-orthorhombic transition is suppressed due to strain, breaking $PT$ and $tT$ symmetry and making strained $Mn_5Si_3$ thin film an AM.[37,38,49]

Experimentally, 100 nm $Mn_5Si_3$(0001) thin film grown on the $Al_2O_3$(0001) substrate follows the epitaxial relationship of $Al_2O_3$(0001)[$01\bar{1}0$]//$Mn_5Si_3$(0001)[$11\bar{2}0$], and undergoes PM-to-AM and AM-to-ncAFM magnetic phase transitions at 243 K and 63 K, respectively (Methods, Extended Data Fig. 1). At weak-fields (up to about 1.5 T), AHE and tiny strain-induced spontaneous canted moment $m$ of ~2 emu $cc^{-1}$ are observed, in agreement with previous studies.[37,49] Due to symmetry, $m$ allows first-order linear anomalous Hall responses, e.g., AHE[37,38,43], anomalous Nernst effect[49,50], and magneto-optical Kerr effect.[37] Crucially, it is the Berry curvature determined by $N$ instead of tiny $m$ that dominates these responses under weak magnetic field.

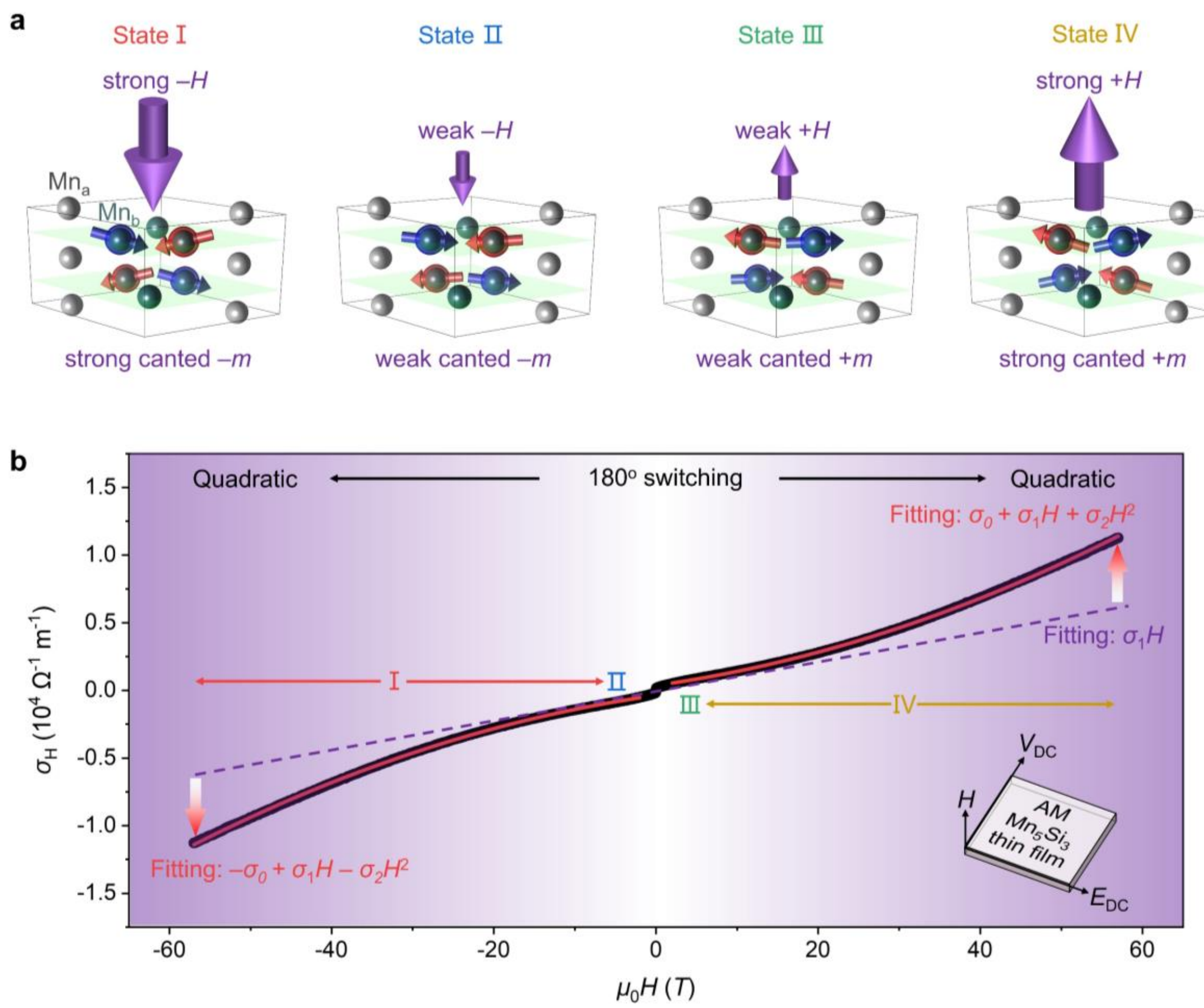

**Figure 2 | Experimental observation of a large magnetic nonlinear Hall effect.** (**a**) Schematic illustrations of the four magnetic states I–IV under strong or weak magnetic field applied in opposite directions. Only Mn atoms of the $Mn_5Si_3$ unit cell are shown for clarity. (**b**) Hall conductivity $\sigma_H$ measured under a perpendicular magnetic field (solid black line) in 100 nm $Mn_5Si_3$(0001) thin film. A gradient background illustrates the smooth evolution from a regime dominated by the AHE due to 180° magnetic structure switching to one characterized by a $\mathrm{sgn}(H)\sigma_2 H^2$ term. The dashed purple and solid red lines denote fitting results using $\sigma_H = \sigma_1 H$ and $\sigma_H = \mathrm{sgn}(H)\sigma_0 + \sigma_1 H + \mathrm{sgn}(H)\sigma_2 H^2$, respectively.

As we discussed in Fig. 1e and 1f, magnetic field $H$ induces Zeeman bands shifts to tune the Berry curvature near the Fermi surface. Next, we discuss the change of order parameter under $H$, which may directly change the Berry curvature itself. Under a weak magnetic field $H$ (~1 T), $m$ can be switched by 180°, simultaneously driving the 180° switching of $N$ (State II and State III in Fig. 2a).[37,49] This 180° switching of $N$ also reverses the spin-splitting bands with CSVL[25], and thereby reversing the first-order linear anomalous Hall responses.[37,49] Using AHE as the readout method, electrical deterministic 180° switching of $N$ has been achieved due to asymmetric energy barriers.[37] In contrast, as shown in State I and State IV of Fig. 2a, under a strong $H$, $m$ increases linearly with $H$ (as discussed in Note S1 in SI and shown in Extended Data Fig. 2), which may further induce $m$-dependent Berry curvature that contributes to higher-order Hall responses with the magnetic field. These four states (I–IV) are experimentally observed, producing a large MNLHE.

As shown in Fig. 2b, the black line depicts the Hall conductivity $\sigma_H$ under strong magnetic field $H$ up to approximately 60 T, measured at 93 K in the AM phase. Notably, this DC $\sigma_H$ can be directly measured under DC longitudinal electric field. Three key features emerge: (1) For State II and State III at weak $H$, a small kink-like offset $\sigma_0$ occurs near zero $H$, corresponding to the weak-field AHE from the 180° switching of $N$.[37] This offset, which does not increase with stronger $H$, is negligible compared to $\sigma_H$ at 60 T, consistent with the alternating-sign Berry curvature at zero $H$. (2) For State I and State IV at stronger $H$, $\sigma_H$ deviates clearly from the OHE-induced linear $\sigma_1 H$, and only a fit including both $\sigma_2 H^2$ and $\sigma_0$ term works. $\sigma_2 H^2$ term in DC measurements proves the existence of large MNLHE. The derivative curve of $\sigma_H$ with respect to $H$ exhibits linear growth under a strong $H$, further confirming the existence of the $\sigma_2 H^2$ term (Extended Data Fig. 3). (3) The $\sigma_H$ curve for positive and negative $H$ cannot be fitted by the same function, because both $\sigma_2 H^2$ and $\sigma_0$ term change sign upon $H$ reversal. This requires two non-analytic sign functions in the fitting function: $\mathrm{sgn}(H)\sigma_2 H^2$ and $\mathrm{sgn}(H)\sigma_0$. This non-analytic feature confirms that both $\mathrm{sgn}(H)\sigma_2 H^2$ and $\mathrm{sgn}(H)\sigma_0$ term are relevant to first-order transitions with $H$ reversal, which is the 180° switching of magnetic structure in our case. The non-analytic feature is quite a distinctive characteristic of MNLHE, which is not shown in traditional NLHE.

MNLHE is reproducible for $Mn_5Si_3$ across the range of temperatures within the AM phase (Fig. 3a).

As the temperature increases from 93 to 200 K, the sgn($H$)$\sigma_2 H^2$ contribution gradually diminishes, which can be attributed to the temperature-sensitive Fermi-Dirac distribution, as discussed later. Meanwhile, sgn($H$)$\sigma_0$ becomes less pronounced at higher temperatures, consistent with a reduction of AHE. In contrast, as shown in Fig. 3b, in the PM phase at 250 K, the sgn($H$)$\sigma_2 H^2$ dependence vanishes entirely, and a perfect linear and analytical dependence of $\sigma_1 H$ is observed. Besides the OHE, the field-linear AHE[51] may also contribute to $\sigma_H$ here, as there might still be some short-range magnetic order or spin fluctuations even in the PM phase[52-54]. This result clearly excludes sgn($H$)$\sigma_2 H^2$ from non-magnetic scenarios (e.g. multi-carrier effects)[44-46] and undoubtably correlates the appearance of sgn($H$)$\sigma_2 H^2$ with the magnetic order in the AM phase.

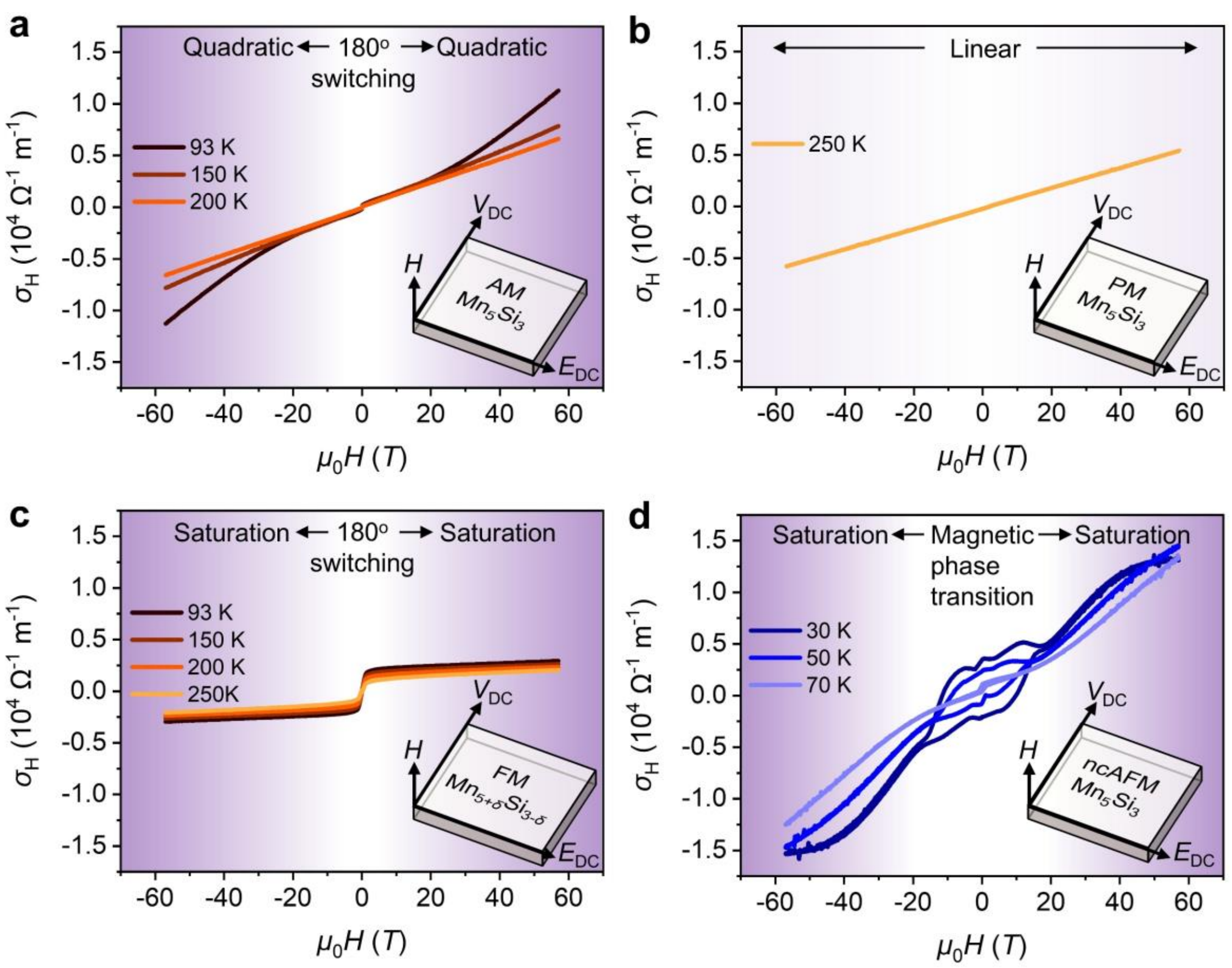

**Figure 3 | Hall conductivity measurements under strong field for $Mn_5Si_3$ in different phases.** Temperature-dependent $\sigma_H$ measured under perpendicular magnetic field for $Mn_5Si_3$(0001) thin film at (**a**) the AM phase, (**b**) the PM phase, (**c**) the FM phase, and (**d**) the ncAFM phase.

To further explore whether this sgn($H$)$\sigma_H(H^2)$ term simply arises from an enhanced FM moment under strong $H$, we carried out control experiments on $Mn_{5+\delta}Si_{3-\delta}$ in the FM phase. The FM phase of $Mn_{5+\delta}Si_{3-\delta}$ is realized by growing thin film at room temperature, which lacks crystallinity while exhibits strong magnetism (Extended Data Fig. 4). As presented in Fig. 3c, besides sgn($H$)$\sigma_0$ induced by 180° FM switching, $\sigma_H$ saturates under strong field for $Mn_{5+\delta}Si_{3-\delta}$ thin film. This result confirms that the sgn($H$)$\sigma_H(H^2)$ observed in the AM phase of $Mn_5Si_3$ thin film is not simply induced by a $H$-enhanced FM moment.

Next, we prove that both Néel vector $N$ and a tiny spontaneous canted moment $m$ under zero field is crucial to generate this sgn($H$)$\sigma_H(H^2)$ dependence. (1) To investigate the role of $N$, we carried out measurements in the ncAFM phase of $Mn_5Si_3$ thin film, as presented in Fig. 3d. The AHE-like hysteresis up to about 20 T is due to the magnetic phase transition triggered by $H$, consistent with previous works[55,56] and beyond the scope of this work. Notably, under strong field, $\sigma_H$ grows linearly with $H$ for 70 K and 50 K, and even tends toward saturation at 30 K. It is clear that these behaviors in the ncAFM phase are distinct from those in the AM phase. This shows that AFM exchange coupling in the ncAFM phase is not enough to bring about the sgn($H$)$\sigma_2 H^2$ dependence. The Néel vector $N$ and corresponding specific magnetic structure in the AM phase plays an important role. (2) Secondly, the tiny spontaneous $m$ is essential. It provides the necessary handle that allows the magnetic structure to be reversed at weak fields. This simultaneously reverses the hopping chirality (as discussed later) and giving rise to the non-analytic characteristic of MNLHE. In contrast, if spontaneous $m$ is strictly zero under zero $H$, even though a strong $H$ can tilt $N$, $N$ will not undergo a 180° switching. Consequently, the magnetic structure cannot be switched by 180°, which cannot lead to the non-analytic characteristic.

The exclusive appearance of non-analytic MNLHE in the AM phase of $Mn_5Si_3$ is reproducible, with the results in another $Mn_5Si_3$ thin film sample presented in Extended Data Fig. 5 as an example. Besides, the non-analytic MNLHE relies on the perpendicular alignment of the magnetic field $H$, electric field $E_{DC}$, and Hall voltage $V_{DC}$, as the $\sigma_H$ is negligible when $H$ is parallel to $E_{DC}$ (Extended

Data Fig. 6). As a result, supported by these comprehensive and consistent data provided above, we can confidently conclude that the non-analytic MNLHE in $Mn_5Si_3$ thin film is intrinsically originated in the response of the magnetic structure of AM under strong $H$.

**Chiral next-nearest-neighbor hopping mechanism for MNLHE**

We move on to explore the mechanism behind this non-analytic MNLHE. The key points are as follows. A tight-binding model of AM $Mn_5Si_3$ is built to derive the Hamiltonian, linking the change in band structure to $H$ (detailed in Note S2 in SI). Next-nearest-neighbor (NNN) hopping through magnetic atoms acquires additional exchange energy proportional to $m$, while also picks up a Haldane-like chiral flux phase[57] proportional to $m$. Together, these factors yield an $m^2$ term, ultimately producing the observed $H^2$ dependence in $\sigma_H$, as determined by the Berry curvature of the band structure (detailed in Note S3 in SI). Importantly, the magnetic structure can be 180° reversed by $H$, which switches the spin-splitting band and flips the hopping chirality, bringing about the non-analytic behavior. The details are below.

For simplicity without losing genericity, we build a 2D hexagonal lattice to represent $Mn_5Si_3$ unit cell, as shown in Fig. 4a and Extended Data Fig. 7. A and B sites are located at the vertices of this lattice, formed by averaging the centroids of all non-magnetic atoms, including both Si and some Mn atoms. Four magnetic sites lie within this 2D plane, with the Néel vector $N$ oriented along the $y$ axis. In this model, the magnetic sublattices with antiferromagnetic coupling are connected by $C_{2x}$ or $C_{2y}T$ symmetry, denoting twofold rotational symmetry along $x$ axis as well as the joint operation of twofold rotational symmetry along $y$ axis and time-reversal symmetry. These symmetries ensuring alternating-sign Berry curvatures at zero field. Meanwhile, $PT$ and $tT$ symmetry are broken. These are consistent with the magnetic structure of AM $Mn_5Si_3$ thin film. Hence, this model captures the essential symmetries of altermagnet $Mn_5Si_3$ and substantially simplifies the Hamiltonian.

Similar to the Haldane model for graphene[57] (Extended Data Fig. 8), both nearest-neighbor (NN)

hopping $t_1$ and NNN hopping $t_2$ are considered, as denoted by black vectors in Fig. 4a. To ensure the final Hamiltonian analytically solvable, we focus on hopping between non-magnetic A and B sites, consistent with the two-band model of the band structure in AM.[31] This assumption is reasonable, as A and B sites account for most of the mass in the $Mn_5Si_3$ unit cell.

We then figure out all these NN and NNN hopping terms. The strong magnetic field $H$ appears as an onsite energy term $-H\sigma_z$. For NN hopping between A and B, no magnetic atoms are involved, so the term simplifies to $t_1\sigma_0$. By contrast, for NNN hopping, the hopping paths encounter magnetic atoms (Fig. 4a), introducing exchange energy from the magnetic atoms. Hence, the NNN hopping term in the Hamiltonian is written as $-t_2\sum_i(\mathbf{m}_i\cdot\boldsymbol{\sigma})$. Nevertheless, this term is proportional to $m$. It cannot explain the $H^2$ dependence of $\sigma_H$, which corresponds to a $m^2$ term.

Inspired by the Haldane model,[57] we consider the additional phase acquired during NNN hopping process, which may contribute to higher-order effects. Specifically, when an electron hops through a magnetic atom, spin-orbit coupling introduces an extra phase $\Phi$. By combining conjugate relations, $P$ symmetry, and $M_yT$ symmetry (as discussed in Note S2 in SI), there are only two independent phases, $\Phi_1$ and $\Phi_2$, corresponding to hopping paths passing through two and one magnetic atom, respectively (Fig. 4b). We can prove that $\Phi_1$ and $\Phi_2$ are odd functions of $m$ by analyzing time-reversal operation and $M_z$ operation, meaning that, in the first-order approximation, $\Phi_1$ and $\Phi_2$ are linearly related to $m$ (as discussed in Note S2 in SI). Therefore, $m^2$ term emerges after considering both of this Haldane-like phase and the exchange energy as we introduced above, modifying $-t_2\sum_i(\mathbf{m}_i\cdot\boldsymbol{\sigma})$ to $-t_2e^{i\Phi}\sum_i(\mathbf{m}_i\cdot\boldsymbol{\sigma})$.

Interestingly, under a strong positive $H$, the positive $m$ constrains these phases in a clockwise hopping direction, forming chiral flux phases (Fig. 4b). Conversely, under a strong negative $H$, the negative $m$ limits these phases in a counter-clockwise hopping chirality (Fig. 4c). It means that, the hopping chirality can be reversed by $H$, due to the 180° switching of $m$. This chiral hopping and its chirality reversal by $H$ induce the non-analytic property of $H^2$ term in $\sigma_H$.

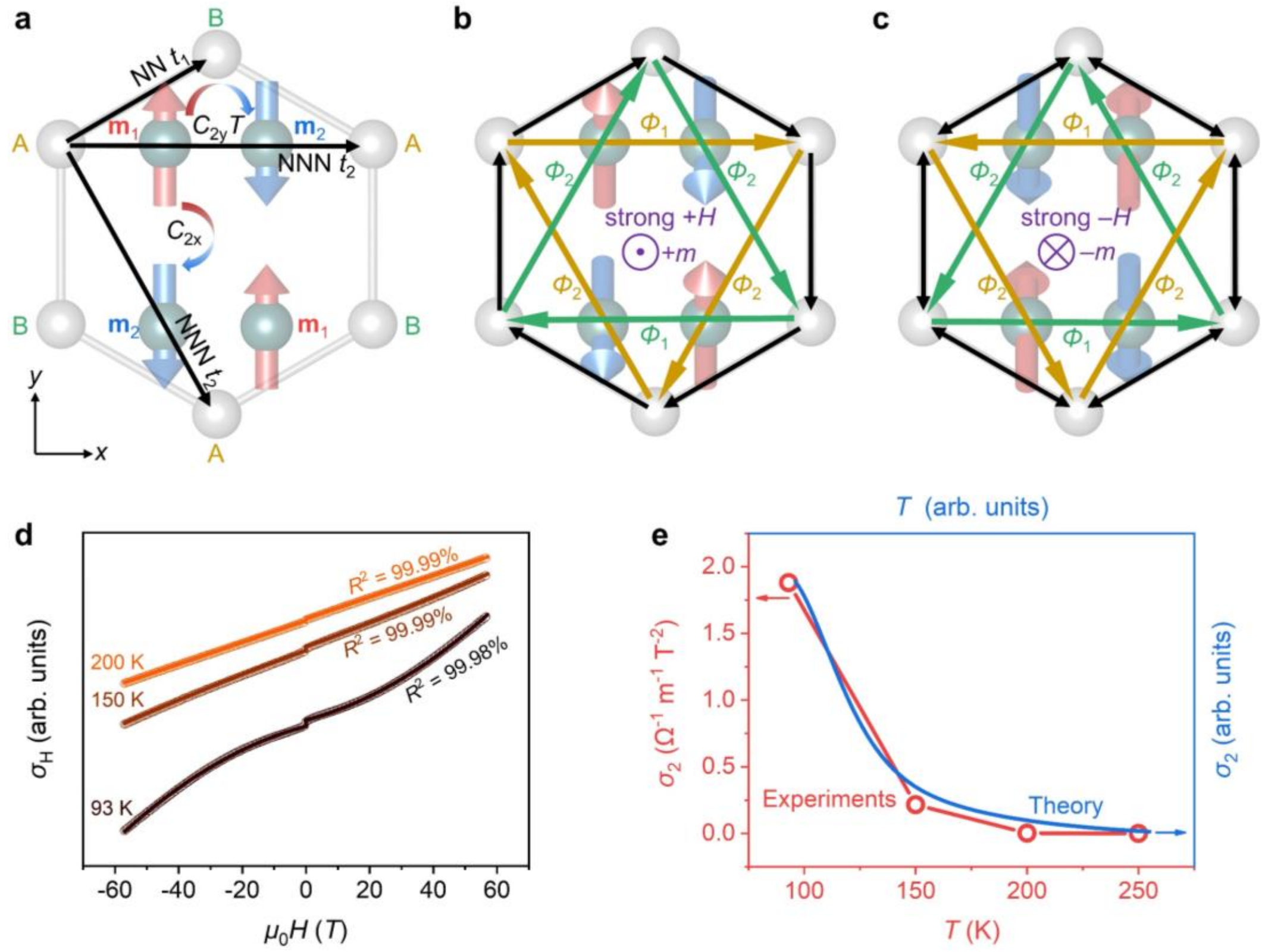

**Figure 4 | Chiral next-nearest-neighbor hopping mechanism for the MNLHE.** (**a**) 2D hexagonal lattice to model $Mn_5Si_3$ unit cell. The magnetic structure and Haldane-like next-nearest-neighbor hopping phases of $Mn_5Si_3$ with (**b**) positive $m$ under positive strong $H$, and (**c**) negative $m$ under negative strong $H$. (**d**) Temperature dependent fitting results of experimental anomalous Hall conductivity $\sigma_H$ data by the theoretical model. $R^2$ is the coefficient of determination. (**e**) The experimental temperature dependence of the quadratic term coefficient $\sigma_2$ (left) and theoretical temperature dependence of $\sigma_2$ (right).

Finally, after some algebra analysis (refer to Note S2 in SI for detailed deviation), the Hamiltonian is as follows, which forms the 2×2 Hamiltonian of $\begin{bmatrix} \mathcal{H}_{K_\pm}^{\uparrow} & 0 \\ 0 & \mathcal{H}_{K_\pm}^{\downarrow} \end{bmatrix}$ presented below.

$$\mathcal{H}_{K_\pm}^{\uparrow} = -H\tau_0 \pm k_x\tau_x + k_y\tau_y \pm m^2\tau_z \tag{1}$$

$$\mathcal{H}_{K_\pm}^{\downarrow} = +H\tau_0 \pm k_x\tau_x + k_y\tau_y \mp m^2\tau_z \tag{2}$$

Remarkably, Eq. (1-2) is very similar with the Hamiltonian of graphene, except for the $m^2\tau_z$ term. This $m^2\tau_z$ term opens the gap of the Dirac cones for the emergence of additional Berry curvature under magnetic field, which gives the final form of Hall conductivity as following Eq. (3):

$$\sigma_H \cong \mathrm{sgn}(H)\,\sigma_0 + \sigma_1 H + \mathrm{sgn}(H)\,\sigma_2 H^2 \tag{3}$$

The sign functions in the Eq. (3) bring about the non-analytic feature in MNLHE. The 180° switching of $N$ changes the sign of the anomalous Hall conductivity, described by the sgn($H$)$\sigma_0$ term, while, the 180° switching of $m$ reverses the chirality of hopping, contributing to sgn($H$)$\sigma_2 H^2$.

Figure 4d shows the fitting results of our experimental data using Eq. (3). The model agrees well with the experimental measurements, with $R^2$ values exceeding 99.98% across different temperatures. The fitted $\sigma_2$ values, summarized in Fig. 4e, increases with decreasing temperatures. Because the intrinsic anomalous Hall conductivity is originated from the integral of Berry curvatures multiplied by Fermi-Dirac distribution, this temperature dependence of $\sigma_2$ reflects how the Fermi-Dirac distribution influences the band occupation. Considering the influence of temperature on the Fermi-Dirac distribution, this temperature dependence of $\sigma_2$ can also be well captured theoretically (Fig. 4e).

**Discussion**

We discuss the requirements for the non-analytic MNLHE here. Alternating-sign Berry curvature instead of single-sign Berry curvature is vital to prevent the higher-order $H$-dependence from being overshadowed by the first-order effect, thus excluding FMs, as confirmed by control experiments on FM $Mn_{5+\delta}Si_{3-\delta}$ (Fig. 3c). Moreover, the non-analytic feature requires a tiny spontaneous canted moment, which can be switched by 180° upon $H$ reversal to flip the hopping chirality. This excludes bulk collinear AFM $Mn_5Si_3$, as its $PT$ and $tT$ symmetry forbid spontaneous net canted moment. In fact, the specific $H^2$ dependence from the $m^2$ term is due to the exchange interactions and Haldane-like hopping phases, which is dependent on the unique crystal and magnetic structure of $Mn_5Si_3$ thin film in the AM phase. Hence, while other magnets with antiferromagnetically coupled moments and tiny spontaneous canting could exhibit MNLHE, the $H^2$ dependence may not appear. ncAFM $Mn_5Si_3$ is a good example, where its complex non-coplanar magnetic structure[48] influences the detail of magnetic exchange and Haldane-like hopping phases, and thus does not bring about $H^2$ dependence (Fig. 3d). As a result, the MNLHE with non-analytic $H^2$ dependence is a fingerprint of AM $Mn_5Si_3$, which cannot be observed in FM $Mn_{5+\delta}Si_{3-\delta}$, collinear AFM or ncAFM $Mn_5Si_3$.

In summary, a large MNLHE, characterized by a non-analytic $\sigma_2 H^2$ term that does not saturate under 60 T, is experimentally observed via DC Hall measurements and theoretically comprehended by chiral NNN hopping mechanism. MNLHE emerges exclusively in the AM phase, while absence in other phases of $Mn_5Si_3$, thereby excluding possible multicarrier artifacts and highlighting the crucial role of the interplay between $H$ and altermagnetism. The $H^2$ dependence originates from the $m^2$ NNN hopping term in the Hamiltonian: one $m$ from exchange interaction, and the other $m$ from a Haldane-like chiral flux phase. This $m^2$ term opens a gap in the alternating spin-splitting bands of AMs with CSVL, with band shift due to Zeeman effect affecting the alternating-sign Berry curvature near the Fermi surface, which ultimately brings about $H^2$ dependence. The non-analytic feature, unique in MNLHE and absent in electric NLHE, is determined by the hopping chirality, which reverses as $H$ changes sign and flips the magnetic structure and spin-splitting bands. This MNLHE, involving higher-order dependence of $\sigma_H$ with $H$, can in principle occur in any magnetic material with alternating-sign Berry curvature and tiny spontaneous canting —conditions well met by some other AMs like MnTe and CrSb. Our findings on MNLHE not only deepen the fundamental understanding of fingerprints in altermagnet $Mn_5Si_3$ and its alternating spin-splitting band structure, but also hold promise in pulsed high-field sensing technologies, supporting both scientific researches and engineering applications.

## Reference

1 Hall, E. H. On a new action of the magnet on electric currents. *Am. J. Math.* **2**, 287-292 (1879).

2 Klitzing, K. v., Dorda, G. & Pepper, M. New Method for High-Accuracy Determination of the Fine-Structure Constant Based on Quantized Hall Resistance. *Phys. Rev. Lett.* **45**, 494-497 (1980).

3 Tsui, D. C., Stormer, H. L. & Gossard, A. C. Two-Dimensional Magnetotransport in the Extreme Quantum Limit. *Phys. Rev. Lett.* **48**, 1559-1562 (1982).

4 Bernevig, B. A., Hughes, T. L. & Zhang, S.-C. Quantum Spin Hall Effect and Topological Phase Transition in HgTe Quantum Wells. *Science* **314**, 1757-1761 (2006).

5 Chang, C.-Z. *et al.* Experimental Observation of the Quantum Anomalous Hall Effect in a Magnetic Topological Insulator. *Science* **340**, 167-170 (2013).

6 Kato, Y. K., Myers, R. C., Gossard, A. C. & Awschalom, D. D. Observation of the Spin Hall Effect in Semiconductors. *Science* **306**, 1910-1913 (2004).

7 Nagaosa, N., Sinova, J., Onoda, S., MacDonald, A. H. & Ong, N. P. Anomalous Hall effect. *Rev. Mod. Phys.* **82**, 1539-1592 (2010).

8 Sodemann, I. & Fu, L. Quantum Nonlinear Hall Effect Induced by Berry Curvature Dipole in Time-Reversal Invariant Materials. *Phys. Rev. Lett.* **115**, 216806 (2015).

9 Ma, Q. *et al.* Observation of the nonlinear Hall effect under time-reversal-symmetric conditions. *Nature* **565**, 337-342 (2019).

10 Du, Z. Z., Lu, H.-Z. & Xie, X. C. Nonlinear Hall effects. *Nat. Rev. Phys.* **3**, 744-752 (2021).

11 Kumar, D. *et al.* Room-temperature nonlinear Hall effect and wireless radiofrequency rectification in Weyl semimetal $TaIrTe_4$. *Nat. Nanotechnol.* **16**, 421-425 (2021).

12 Kang, K., Li, T., Sohn, E., Shan, J. & Mak, K. F. Nonlinear anomalous Hall effect in few-layer $WTe_2$. *Nat. Mater.* **18**, 324-328 (2019).

13 Xiao, J. *et al.* Berry curvature memory through electrically driven stacking transitions. *Nat. Phys.* **16**, 1028-1034 (2020).

14 Ho, S.-C. *et al.* Hall effects in artificially corrugated bilayer graphene without breaking time-reversal symmetry. *Nat. Electron.* **4**, 116-125 (2021).

15 Lai, S. *et al.* Third-order nonlinear Hall effect induced by the Berry-connection polarizability tensor. *Nat. Nanotechnol.* **16**, 869-873 (2021).

16 Gao, A. Y. *et al.* Quantum metric nonlinear Hall effect in a topological antiferromagnetic heterostructure. *Science* **381**, 181-186 (2023).

17 Wang, N. *et al.* Quantum-metric-induced nonlinear transport in a topological antiferromagnet. *Nature* **621**, 487-492 (2023).

18 He, P. *et al.* Third-order nonlinear Hall effect in a quantum Hall system. *Nat. Nanotechnol.* **19**, 1460-1465 (2024).

19 He, Z. & Weng, H. Giant nonlinear Hall effect in twisted bilayer $WTe_2$. *npj Quantum Mater.* **6**, 101 (2021).

20 Huang, M. Z. *et al.* Intrinsic Nonlinear Hall Effect and Gate-Switchable Berry Curvature Sliding in Twisted Bilayer Graphene. *Phys. Rev. Lett.* **131**, 066301 (2023).

21 Zhao, T.-Y. *et al.* Gate-Tunable Berry Curvature Dipole Polarizability in Dirac Semimetal $Cd_3As_2$. *Phys. Rev. Lett.* **131**, 186302 (2023).

22 Min, L. *et al.* Strong room-temperature bulk nonlinear Hall effect in a spin-valley locked Dirac material. *Nat. Commun.* **14**, 364 (2023).

23 Lee, J.-E. *et al.* Spin-orbit-splitting-driven nonlinear Hall effect in $NbIrTe_4$. *Nat. Commun.* **15**, 3971 (2024).

24 Han, J. *et al.* Room-temperature flexible manipulation of the quantum-metric structure in a topological chiral antiferromagnet. *Nat. Phys.* **20**, 1110-1117 (2024).

25 Ma, H. Y. *et al.* Multifunctional antiferromagnetic materials with giant piezomagnetism and noncollinear spin current. *Nat. Commun.* **12**, 2846 (2021).

26 Zhang, C.-P., Gao, X.-J., Xie, Y.-M., Po, H. C. & Law, K. T. Higher-order nonlinear anomalous Hall effects induced by Berry curvature multipoles. *Phys. Rev. B* **107**, 115142 (2023).

27 Sankar, S. *et al.* Experimental Evidence for a Berry Curvature Quadrupole in an Antiferromagnet. *Phys. Rev. X* **14**, 021046 (2024).

28 He, P. *et al.* Bilinear magnetoelectric resistance as a probe of three-dimensional spin texture in topological surface states. *Nat. Phys.* **14**, 495-499 (2018).

29 Kim, D.-J. *et al.* Spin Hall-induced bilinear magnetoelectric resistance. *Nat. Mater.* **23**, 1509-1514 (2024).

30 Šmejkal, L., Sinova, J. & Jungwirth, T. Beyond Conventional Ferromagnetism and Antiferromagnetism: A Phase with Nonrelativistic Spin and Crystal Rotation Symmetry. *Phys. Rev. X* **12**, 031042 (2022).

31 Šmejkal, L., Sinova, J. & Jungwirth, T. Emerging Research Landscape of Altermagnetism. *Phys. Rev. X* **12**, 040501 (2022).

32 Song, C. *et al.* Altermagnets as a new class of functional materials. *Nat. Rev. Mater.*, doi:10.1038/s41578-025-00779-1 (2025).

33 Amin, O. J. *et al.* Nanoscale imaging and control of altermagnetism in MnTe. *Nature* **636**, 348-353 (2024).

34 Krempaský, J. *et al.* Altermagnetic lifting of Kramers spin degeneracy. *Nature* **626**, 517-522 (2024).

35 Feng, Z. *et al.* An anomalous Hall effect in altermagnetic ruthenium dioxide. *Nat. Electron.* **5**, 735-743 (2022).

36 Gonzalez Betancourt, R. D. *et al.* Spontaneous Anomalous Hall Effect Arising from an Unconventional Compensated Magnetic Phase in a Semiconductor. *Phys. Rev. Lett.* **130**, 036702 (2023).

37 Han, L. *et al.* Electrical 180° switching of Néel vector in spin-splitting antiferromagnet. *Sci. Adv.* **10**, eadn0479 (2024).

38 Reichlova, H. *et al.* Observation of a spontaneous anomalous Hall response in the $Mn_5Si_3$ d-wave altermagnet candidate. *Nat. Commun.* **15**, 4961 (2024).

39 Kluczyk, K. P. *et al.* Coexistence of anomalous Hall effect and weak magnetization in a nominally collinear antiferromagnet MnTe. *Phys. Rev. B* **110**, 155201 (2024).

40 Zhou, Z. *et al.* Manipulation of the altermagnetic order in CrSb via crystal symmetry. *Nature* **638**, 645-650 (2025).

41 Cheong, S.-W. & Huang, F.-T. Altermagnetism with non-collinear spins. *npj Quantum Mater.* **9**, 13 (2024).

42 Fang, Y., Cano, J. & Ghorashi, S. A. A. Quantum Geometry Induced Nonlinear Transport in Altermagnets. *Phys. Rev. Lett.* **133**, 106701 (2024).

43 Leiviskä, M. *et al.* Anisotropy of the anomalous Hall effect in thin films of the altermagnet candidate $Mn_5Si_3$. *Phys. Rev. B* **109**, 224430 (2024).
44 Urata, T., Hattori, W. & Ikuta, H. High mobility charge transport in a multicarrier altermagnet CrSb. *Phys Rev Mater* **8**, 084412 (2024).
45 Bai, Y. *et al.* Nonlinear field dependence of Hall effect and high-mobility multi-carrier transport in an altermagnet CrSb. *Appl Phys Lett* **126**, 042402 (2025).
46 Peng, X. *et al.* Scaling Behavior of Magnetoresistance and Hall Resistivity in Altermagnet CrSb. *arXiv: 2412.12263* (2024).
47 Brown, P. & Forsyth, J. Antiferromagnetism in $Mn_5Si_3$: the magnetic structure of the AF2 phase at 70 K. *J Phys Condens Matter* **7**, 7619 (1995).
48 Surgers, C., Fischer, G., Winkel, P. & Lohneysen, H. V. Large topological Hall effect in the non-collinear phase of an antiferromagnet. *Nat Commun* **5**, 3400 (2014).
49 Han, L. *et al.* Observation of non-volatile anomalous Nernst effect in altermagnet with collinear Néel vector. *arXiv:2403.13427* (2024).
50 Badura, A. *et al.* Observation of the anomalous Nernst effect in altermagnetic candidate $Mn_5Si_3$. *arXiv:2403.12929* (2024).
51 Li, X., Koo, J., Zhu, Z., Behnia, K. & Yan, B. Field-linear anomalous Hall effect and Berry curvature induced by spin chirality in the kagome antiferromagnet $Mn_3Sn$. *Nat. Commun.* **14**, 1642 (2023).
52 Gerashenko, A. *et al.* Short-range magnetic order in the paramagnetic phase of cubic $SrMnO_{3-x}$ ($x<0.005$): An $^{17}O$ and $^{87}Sr$ NMR study. *Phys. Rev. B* **102**, 134408 (2020).
53 Gray, I. *et al.* Time-resolved magneto-optical effects in the altermagnet candidate MnTe. *Appl. Phys. Lett.* **125**, 212404 (2024).
54 Wei, C.-C. *et al.* $La_2O_3Mn_2Se_2$: A correlated insulating layered *d*-wave altermagnet. *Phys. Rev. Mater.* **9**, 024402 (2025).
55 Sürgers, C. *et al.* Switching of a large anomalous Hall effect between metamagnetic phases of a non-collinear antiferromagnet. *Sci. Rep.* **7**, 42982 (2017).
56 Sürgers, C., Kittler, W., Wolf, T. & Löhneysen, H. v. Anomalous Hall effect in the noncollinear antiferromagnet $Mn_5Si_3$. *AIP Adv.* **6**, 055604 (2016).
57 Haldane, F. D. M. Model for a Quantum Hall Effect without Landau Levels: Condensed-Matter Realization of the "Parity Anomaly". *Phys. Rev. Lett.* **61**, 2015-2018 (1988).
58 Evans, R. F. L. *et al.* Atomistic spin model simulations of magnetic nanomaterials. *J. Phys. Condens. Matter* **26**, 103202 (2014).
59 dos Santos, F. J. *et al.* Spin waves in the collinear antiferromagnetic phase of $Mn_5Si_3$. *Phys. Rev. B* **103**, 024407 (2021).

## Methods

**Sample preparation.** 100 nm $Mn_5Si_3(0001)$ films were grown on $Al_2O_3(0001)$ substrate by co-sputtering of Mn and Si at 600 ºC with a rate of 0.4 Å $s^{-1}$ under a base pressure below $5\times10^{-8}$ Torr. For pulsed high field measurements, 0.15 mm×0.15 mm square Cr(10 nm)/Au(80 nm) electrodes were deposited for connection by Au wire and silver conductive paste. All the samples were kept in a glove box with $O_2$ and $H_2O$ < 0.01 parts per million to prevent degradation or oxidation.

**Characterizations.** Cross-sectional high-angle annular dark-field scanning transmission electron microscopy (HAADF-STEM) was conducted on an FEI Titan 80-300 electron microscopy equipped with a monochromator unit, a probe spherical aberration corrector, a post-column energy filter system (Gatan Tridiem 865 ER), and a Gatan 2k slow scan CCD system, operating at 300 kV, combining an energy resolution of ~0.6 eV and a dispersion of 0.2 eV per channel with a spatial resolution of ~0.08 nm. The longitudinal resistivity and anomalous Hall resistivity of $Mn_5Si_3$ films were measured using commercial Physical Property Measurement System (PPMS, Quantum Design). Magnetic hysteresis curves were collected for 5 mm×3 mm×100 nm $Mn_5Si_3$(0001) thin films on 0.5 mm-thick $Al_2O_3$(0001) substrate under out-of-plane magnetic field, using a commercial superconducting quantum interference device (SQUID, MPMS3), where the diamagnetic contribution of substrate was subtracted.

**Tight binding model calculations.** The tight-binding model was based on the 2D hexagonal lattice with two nonmagnetic sites and four magnetic sites in one cell. The bases were chosen as the two nonmagnetic sites with two spin degrees of freedom, while the magnetic atoms influence the hopping terms. The low-energy approximation was applied that Hamiltonian can be decomposed into two 2×2 matrices. The intrinsic anomalous Hall conductivity can be analytically calculated from the integral of Berry curvatures on occupied states.

**Atomic spin simulations.** Atomic spin simulations were carried out on VAMPIRE software,[58] which can define specific lattice parameters $a$ and $c$ consistent with our experimental values for $Mn_5Si_3$ thin film.[49] Discretized 10×10×5 units cells in the $x$-$y$-$z$ direction were used to model the $Mn_5Si_3$ thin. Four exchange interactions were considered, namely $J_1$ = –12.23 meV, $J_2$ = –2.16 meV, $J_3$ = 3.98 meV, and $J_4$ = 2.89 meV, with magnitude consistent with former DFT calculations.[37,59] The DMI vector was set along $x$ axis with a magnitude of 1.35 meV.[37]

**Data availability**. All data needed to evaluate the conclusions in the paper are present in the paper and/or the Supplementary Materials.

**Acknowledgements**. This work was supported by the National Key R&D Program of China (grant nos. 2021YFA1401500 and 2022YFA1402603), National Natural Science Foundation of China (grant nos. 52225106, 12241404), and Hong Kong Research Grants Council (grant nos. 16303821, 16306722, and 16304523).

**Author contributions**. L. H., C. S. and F. P. proposed the ideas and designed the experiments. X. F. and J. L. carried out theoretical analysis and calculations. L. H., Y. Z., and X. L. performed pulsed high field measurements. This work was conceived, led, coordinated and guided by C. S., J. L. and F. P. All the authors contributed to the writing of the manuscript.

**Competing interests**. The authors declare no competing interests.

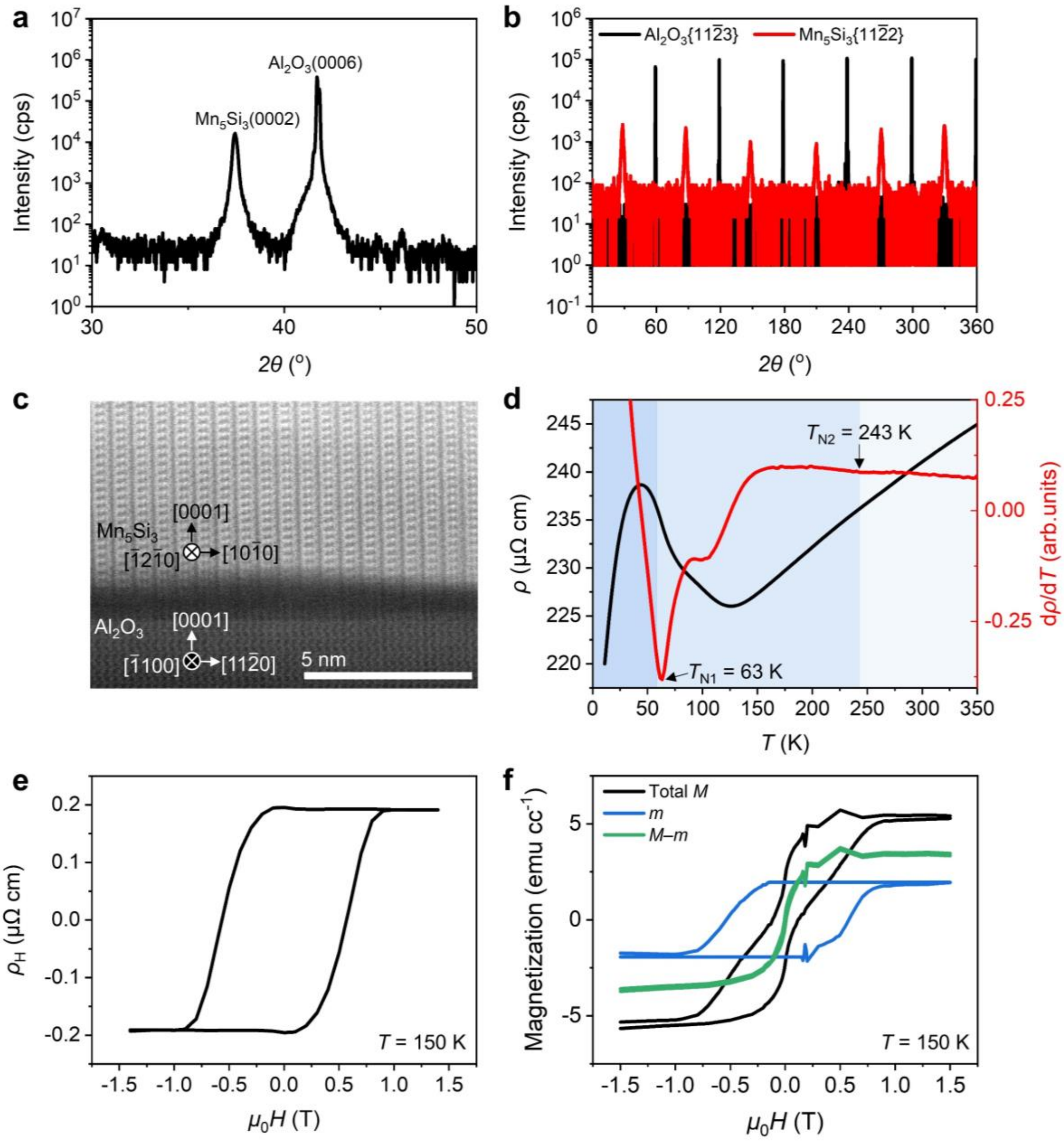

**Extended Data Fig. 1 | Basic characterizations of the $Mn_5Si_3$ thin film.** (**a**) X-ray diffraction pattern of $Mn_5Si_3$(0001) thin film. (**b**) $\varphi$-scan X-ray diffraction measurements for $Al_2O_3$(0001)/$Mn_5Si_3$(0001). An in-plane 30°-rotated growth mode is confirmed from the 30° shift between $Mn_5Si_3[11\bar{2}2]$ and $Al_2O_3[11\bar{2}3]$ peaks. (**c**) HAADF-STEM image of $Al_2O_3$(0001)/$Mn_5Si_3$(0001) under zone axis of $Al_2O_3\langle\bar{1}100\rangle$. (**d**) Temperature-dependent longitudinal resistivity $\rho$ with temperature $T$ (black line) and its temperature derivative (red line). (**e**) Hysteresis of anomalous Hall resistivity $\rho_H$ and (**f**) total magnetization $M$, canted moment $m$ and defect-induced $M$–$m$ measured at 150 K. For $\rho_H$ and $M$, field-linear background was subtracted.

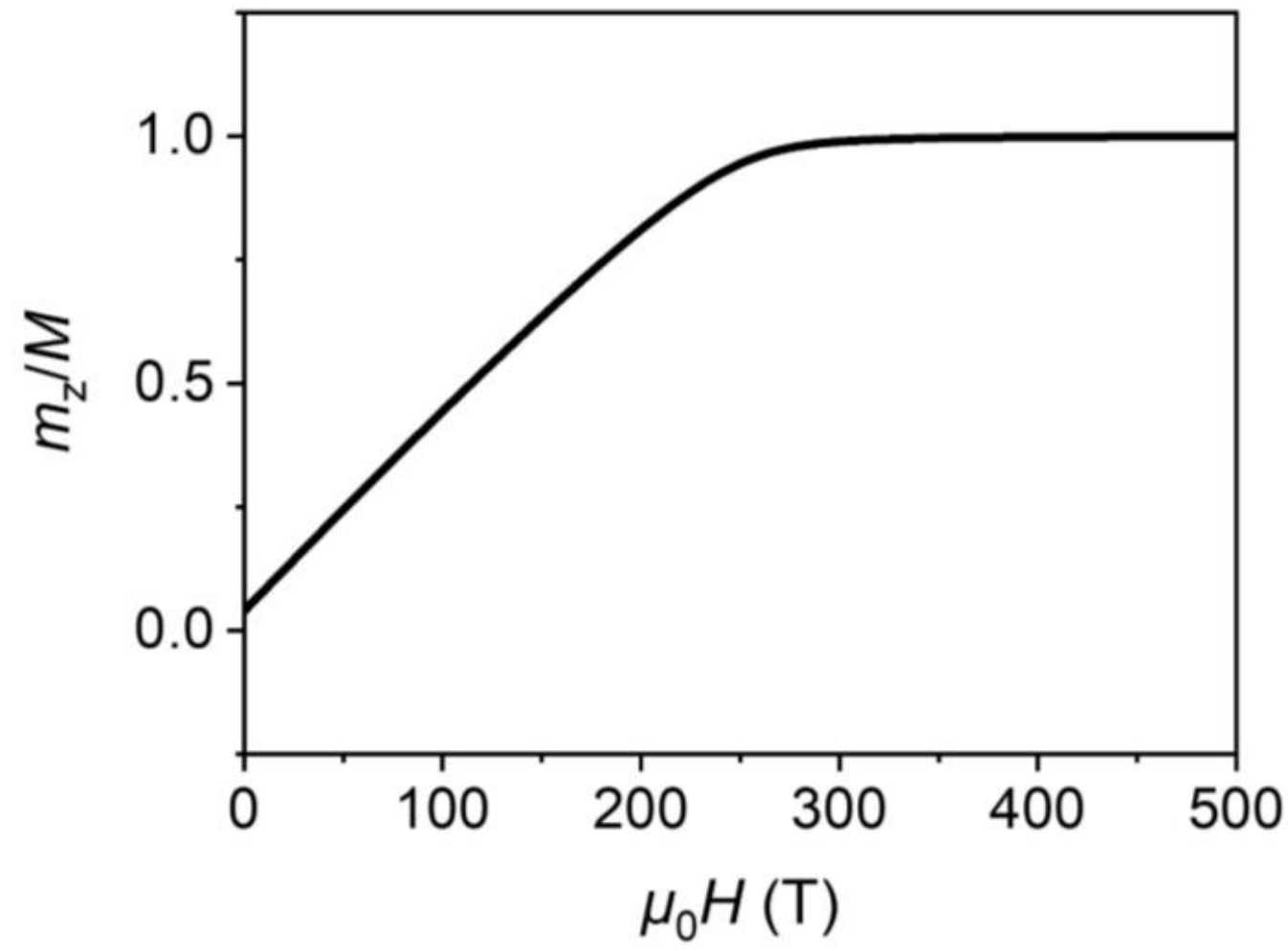

**Extended Data Fig. 2 | Canted *m* along *z* direction under strong out-of-plane magnetic field *H* for AM phase of $Mn_5Si_3$, calculated by atomic spin simulations.**

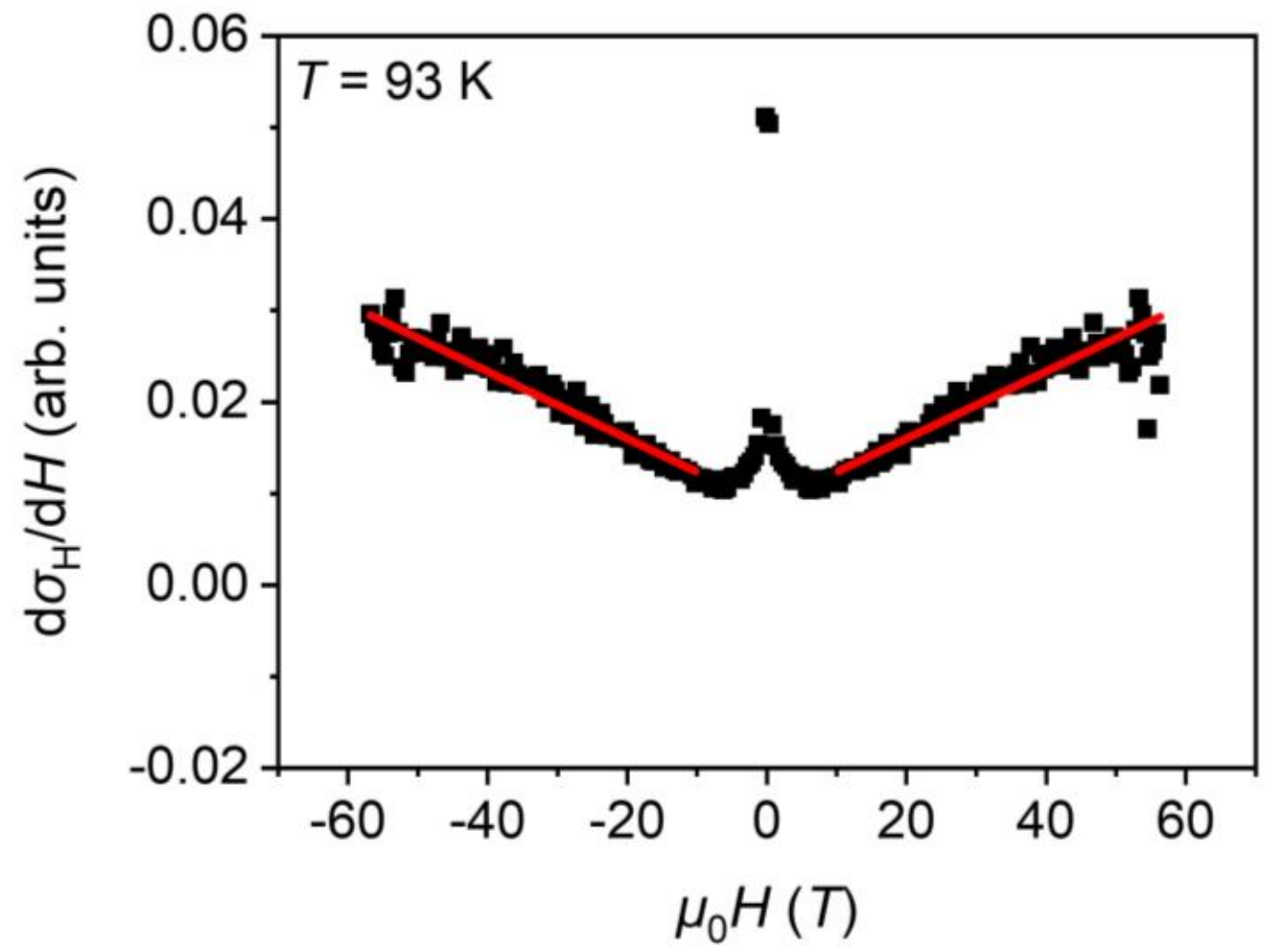

**Extended Data Fig. 3 | The derivative curve of $\sigma_H$ with respect to *H* for Fig. 2b. Red lines show linear fitting at strong field.**

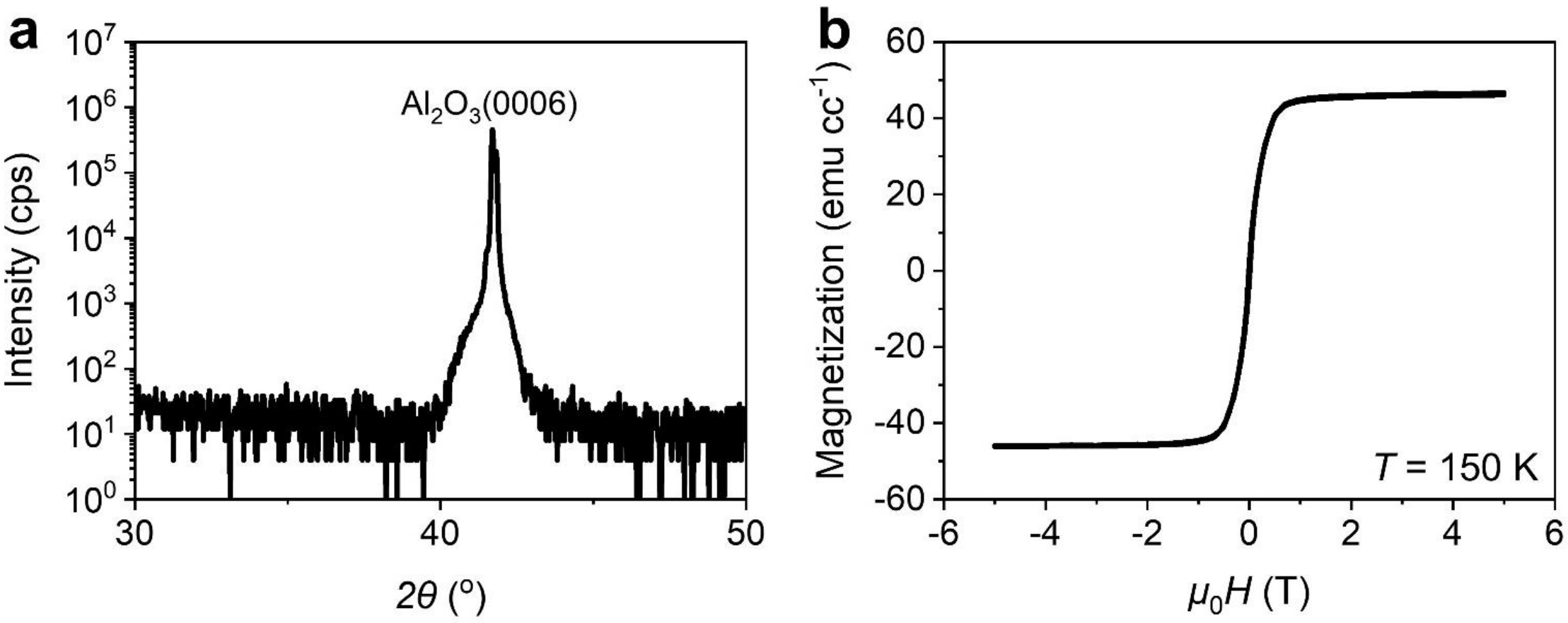

**Extended Data Fig. 4 | Characterizations of ferromagnetic $Mn_{5+\delta}Si_{3-\delta}$ thin film.** (**a**) X-ray diffraction pattern. (**b**) Magnetic hysteresis measured at 150 K. The diamagnetic contribution from the substrate was subtracted.

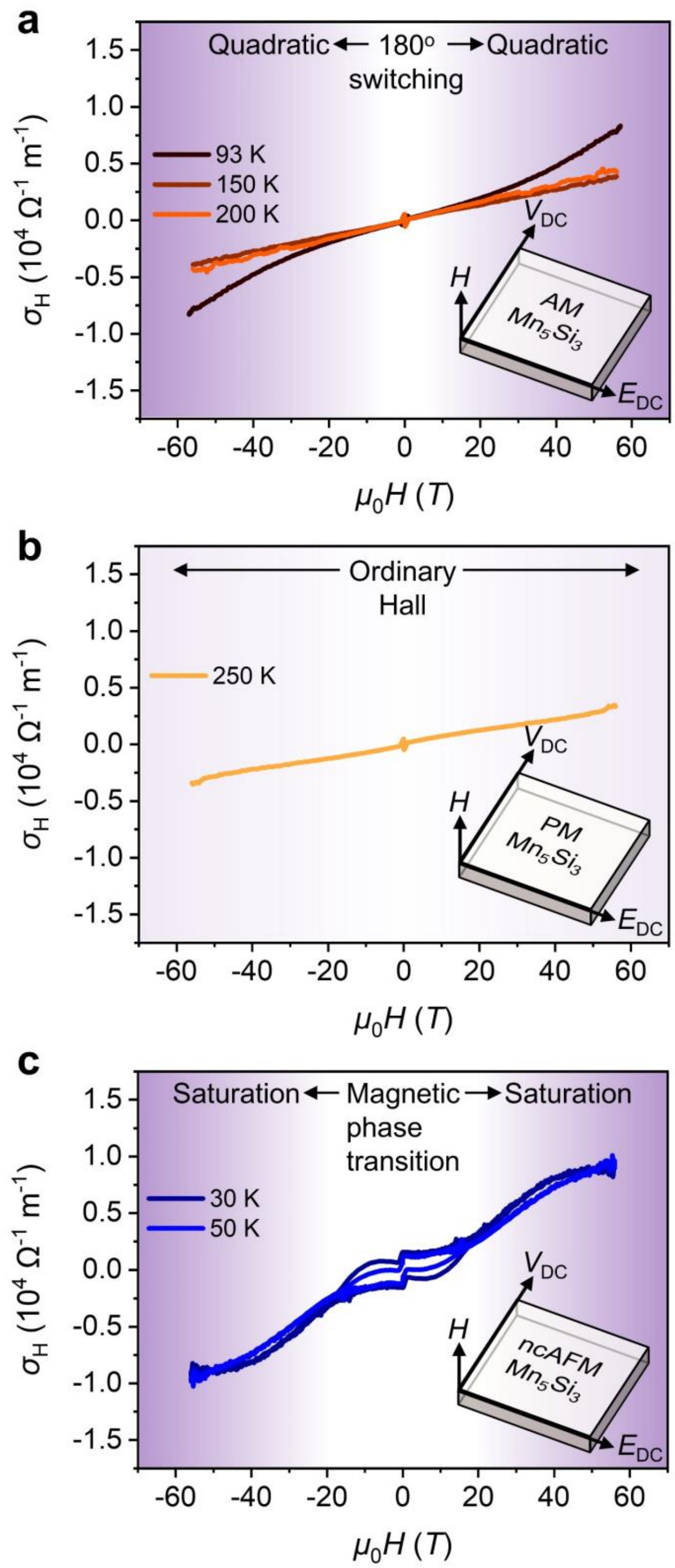

**Extended Data Fig. 5 | Temperature dependent measurements of MNLHE on another $Mn_5Si_3$ thin film sample.** $\sigma_H$ measured under perpendicular magnetic field for $Mn_5Si_3$(0001) thin film at (**a**) the AM phase, (**b**) the PM phase, and (**c**) the ncAFM phase.

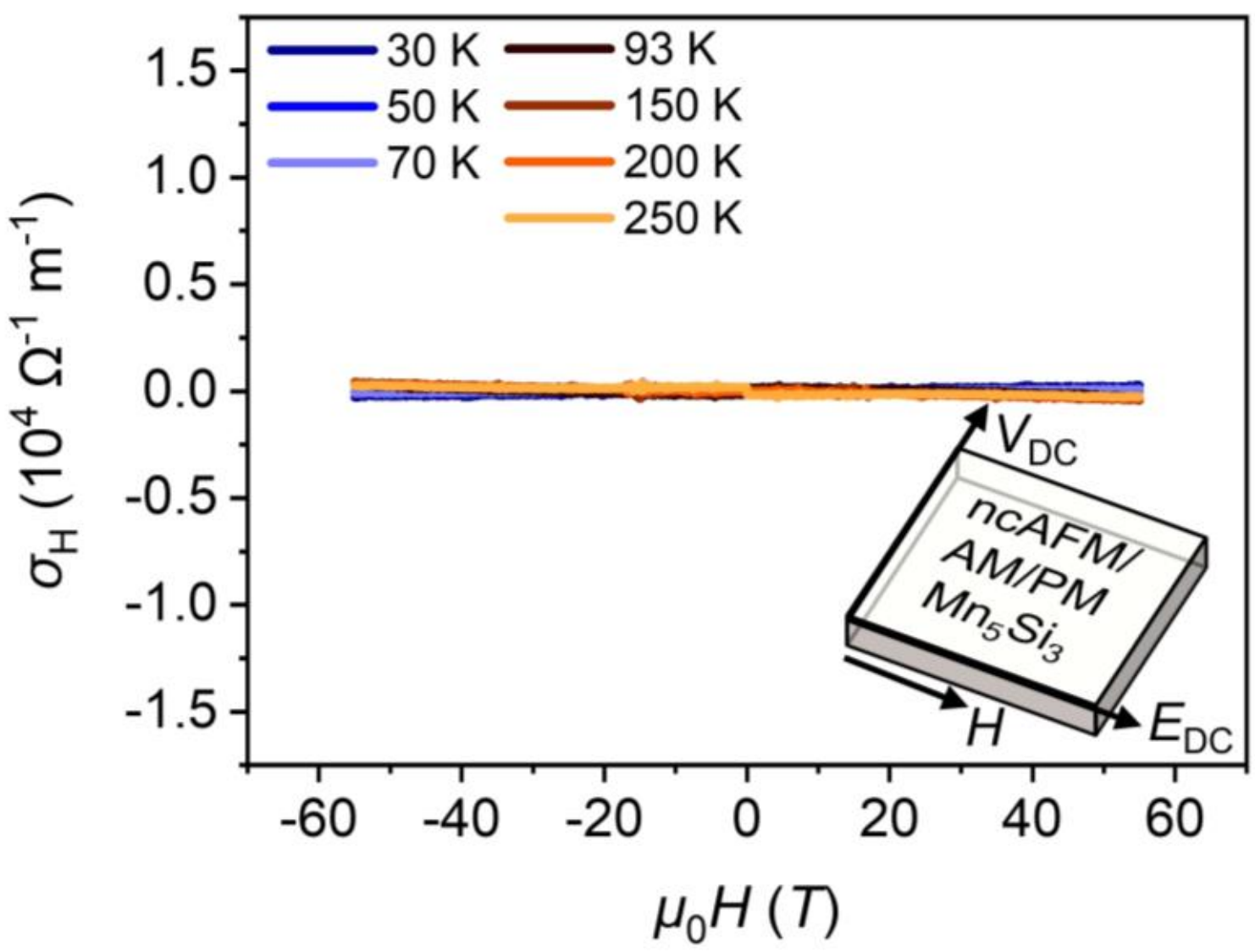

**Extended Data Fig. 6 | Temperature dependent Hall conductivity measured with parallel configuration of magnetic field and electric field.**

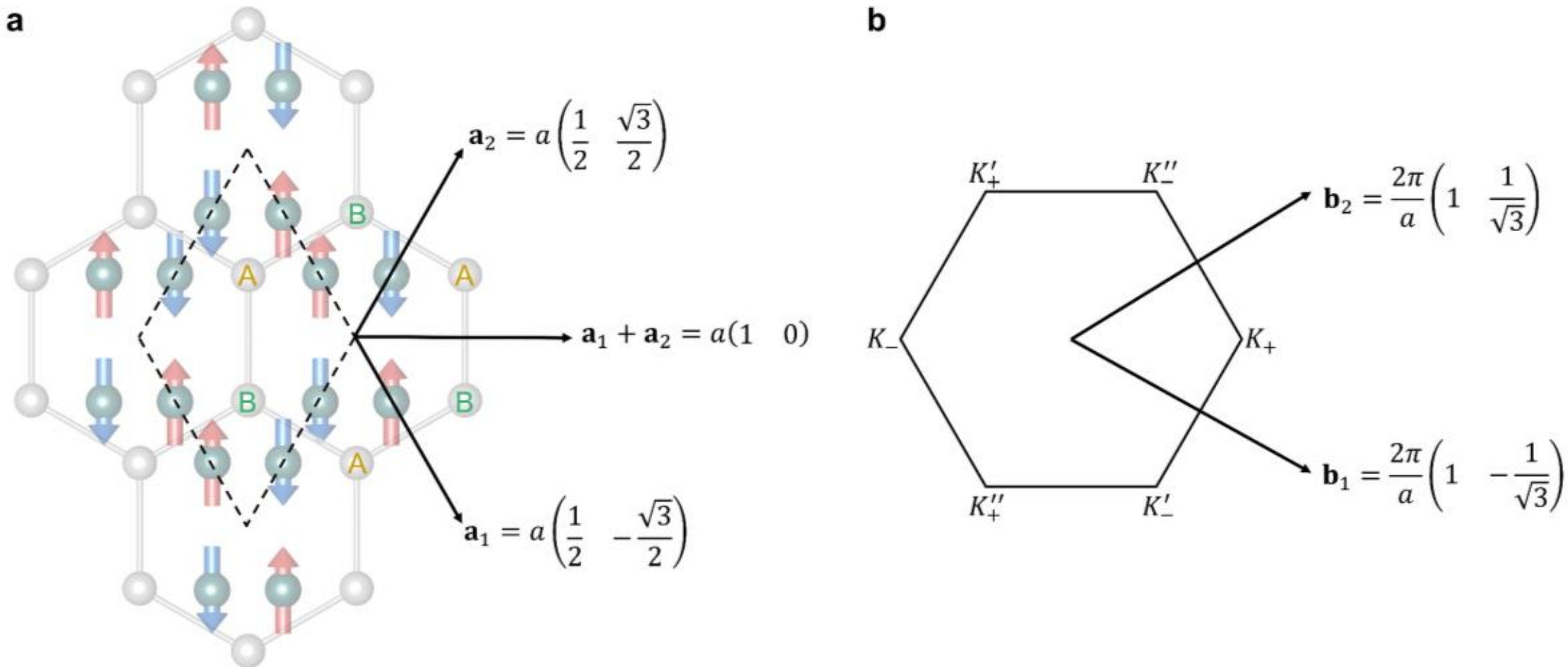

**Extended Data Fig. 7 | Schematic of the 2D hexagonal lattice in (a) real space and (b) reciprocal space.**

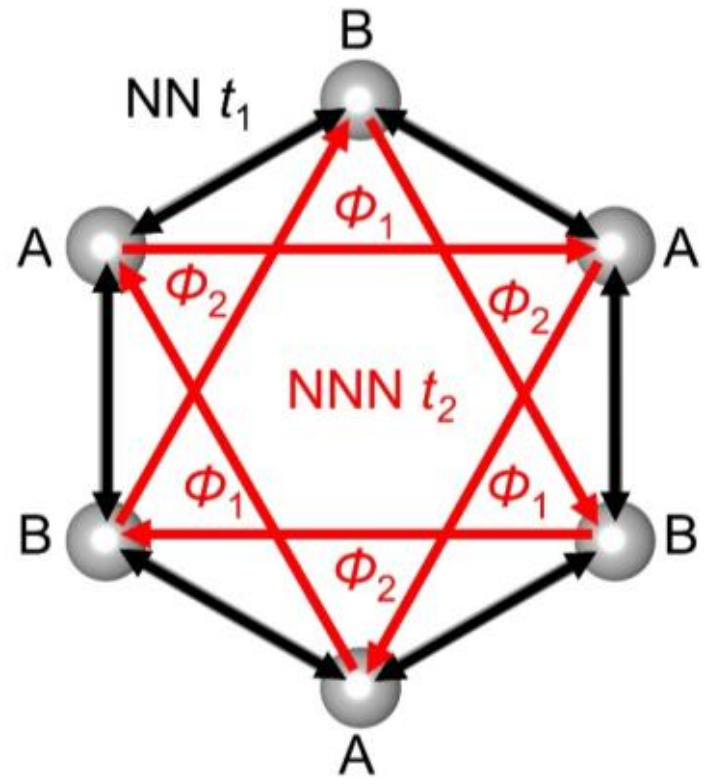

**Extended Data Fig. 8 | Schematic of the Haldane model in graphene.**

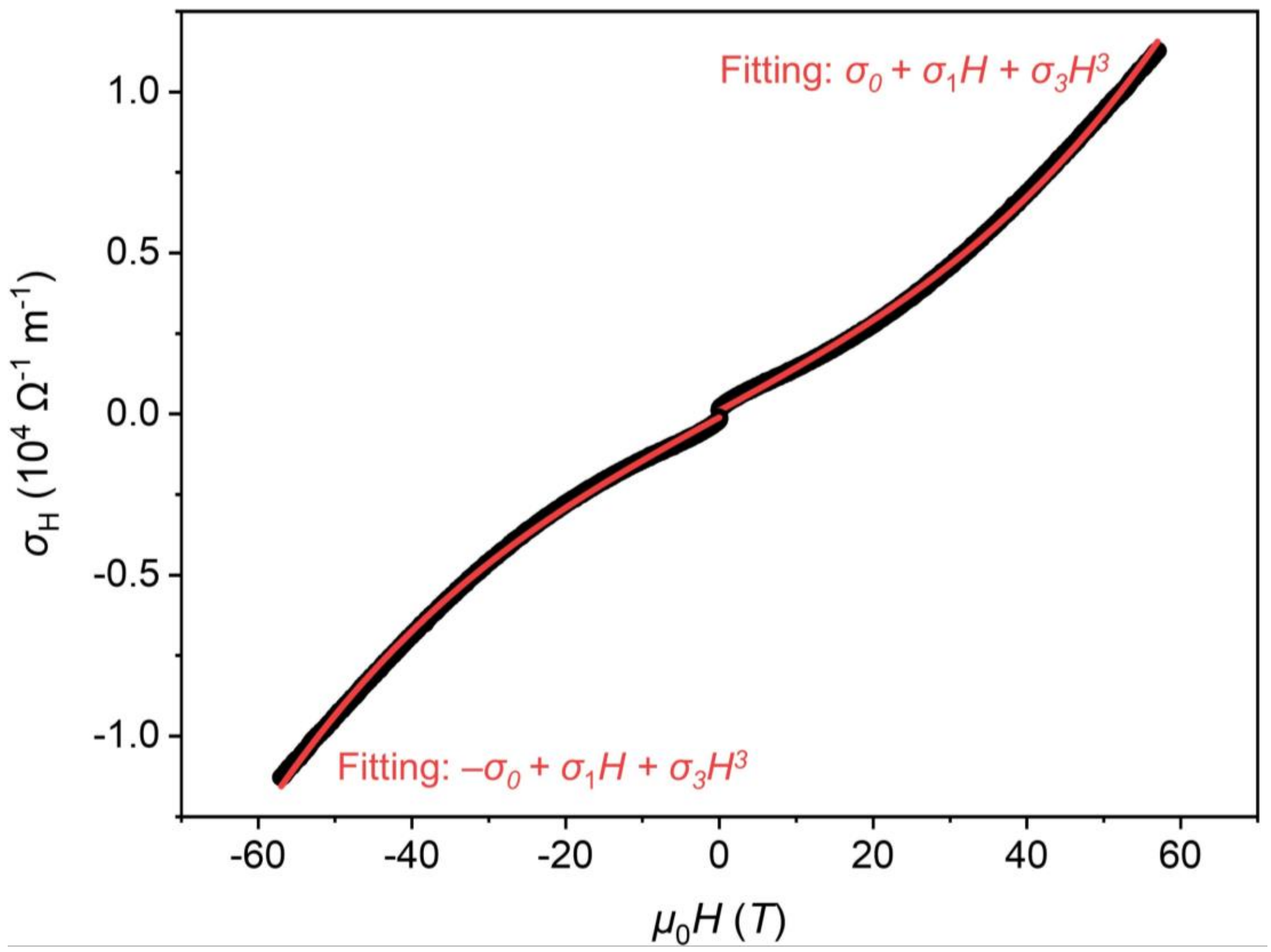

**Extended Data Fig. 9 | Fitting of the Hall conductivity at 93 K by considering an analytic $\sigma_3(H^3)$ term instead of the non-analytic sgn($H$)$\sigma_2 H^2$ term.** It can be seen that the fitted curve differs obviously from the experimental data.